\documentclass[12pt,nofootinbib]{revtex4}
\usepackage{amsmath}
\usepackage{amssymb}
\usepackage{epsfig}
\usepackage{tabularx}
\usepackage{graphicx}
\usepackage{inputenc}
\usepackage{url}
\usepackage{breqn}
\usepackage{hyperref}
\usepackage{footmisc}
\def\no{\tilde{\chi}^0_1}
\def\nt{\tilde{\chi}^0_2}
\def\nth{\tilde{\chi}^0_3}
\def\nf{\tilde{\chi}^0_4}
\def\co{\tilde{\chi}^{\pm}_1}

\def\simlt{\stackrel{<}{{}_\sim}}
\def\simgt{\stackrel{>}{{}_\sim}}

\begin{document}
\begin{titlepage}
\vspace{1.cm}
\title{Blind Spots for neutralino Dark Matter in the MSSM \\ with an intermediate {\large $m_A$}}
\vspace{1.5cm}
\author{\textbf{Peisi Huang$^{a,c}$ and Carlos E.M. Wagner$^{a,b,c}$} \\
\vspace{1.5cm}
\normalsize\emph{$^a$Enrico Fermi Institute \& $^b$Kavli Institute for Cosmological Physics,}\\
\normalsize\emph{University of Chicago, Chicago, IL 60637} \\
\normalsize\emph{$^c$HEP Division, Argonne National Laboratory, 9700 Cass Ave., Argonne, IL 60439}
\vspace{1.5cm}
}

\begin{abstract}
We study the spin-independent neutralino Dark Matter scattering off heavy nuclei in the MSSM. We identify analytically the blind spots in direct detection for intermediate values of $m_A$. In the region where $\mu$ and $M_{1,2}$ have opposite signs, there is not only a reduction of the lightest CP-even Higgs coupling to neutralinos, but also a destructive interference between the neutralino scattering through the exchange of the  lightest CP-even Higgs  and that through the exchange of  the heaviest CP-even  Higgs. At critical values of $m_A$, the tree-level contribution from the light Higgs exchange cancels the contribution from the heavy Higgs, so the scattering cross section vanishes.  We denote these configurations as blind spots, since they provide a generalization of the ones previously discussed in the literature, which occur at very large values of $m_A$.  We show that the generalized blind spots may occur in regions of parameter space  that are consistent with the obtention of the proper neutralino relic density, and can be tested by non-standard Higgs boson searches and EWino searches at the LHC and future linear colliders.   
\end{abstract}
\maketitle
\end{titlepage}
\section{Introduction}
Low energy supersymmetry provides a well motivated extension of the Standard Model (SM),  in which the  weak scale is linked to the scale of supersymmetry (SUSY) breaking.  Minimal supersymmetric extensions of the SM (MSSM) with scalar quarks masses of the order of the TeV and chargino and neutralino masses of the order of the weak scale may be consistent with the recently observed Higgs boson and include a natural Dark Matter (DM) candidate, namely the lightest neutralino~\cite{Nilles:1983ge},\cite{Haber:1984rc},\cite{Martin:1997ns}.   

The LHC collider  collaborations have not  observed any signal of beyond the SM physics yet, and the current bounds on the masses of gluinos and the first and second generation squarks (assuming them to be degenerate in mass)  are now larger than 1 TeV~\cite{ATLAS:colored,CMS:colored}. There are also direct LHC searches on neutralinos and charginos, but the limits are still weak at present~\cite{ATLAS:EW,CMS:EW}.  In the meantime, the direct DM detection experiments (DDMD) are putting limits on the cross section of a weakly interacting massive particle (WIMP) scattering off a target of heavy nuclei. By far, the best limits are from the LUX experiment~\cite{LUX}, which probed the spin independent (SI) elastic scattering, and the Xenon100 Experiment, which set limits on both SI scattering and spin dependent scattering~\cite{Xenon100}.  These experiments, together with the future experiments like Xenon1T~\cite{Xenon1T},  LZ~\cite{LZ} and DarkSide~\cite{DarkSide} are pushing the limits quickly to the regime where the LSP might become visible. The neutralino DM also gets constrained by the relic abundance and indirect DM detection experiments (IDMD), such as IceCube~\cite{IceCube}, Fermi~\cite{FermiLAT}, PAMELA~\cite{Pamela} and AMS02 experiment~\cite{AMS02} .

Many works have tried to identify the MSSM parameter space allowed by current experiments and to understand the prospective for future experiments~\cite{Mandic:2000jz,BirkedalHansen:2001is,BirkedalHansen:2002am,Baer:2005zc,Baer:2005jq,Farina:2011bh,Perelstein:2012qg,Grothaus:2012js,Altmannshofer:2012ks,Hisano:2012wm,Han:2013gba,Hill:2013hoa}. In a recent work~ \cite{Cheung:BS}, the authors have considered a simplified model, where all scalars, including the heavy Higgs bosons and sfermions are decoupled from observational properties of the DM. They pointed out that in the region where the Higgsino mass $\mu$ is negative, the SI scattering cross section can be suppressed for certain values of Bino mass $M_1$, Wino mass $M_2$ and $\tan\beta$.  While the limit of heavy squarks  may be motivated by the current experimental limits on colored particles, unless $\tan\beta$ is very large, there is less motivation to assume very  non-standard Higgs bosons.  In particular, for moderate or large values of $\tan\beta$, preferred to obtain the observed value of the SM-like lightest CP-even Higgs boson mass, the non-standard heavy Higgs bosons may have an important impact on DDMD experimental results.  

In this article, we include the contribution from the heavy CP-even Higgs boson, which is not necessarily decoupled. We will show that there can be constructive or destructive interference between the contribution from a heavy Higgs boson and a light Higgs boson, and the latter one results in new blind spots.  This effect was noticed by some earlier studies~\cite{Mandic:2000jz,Ellis:2000ds,PhysRevD.63.065016,Ellis:2005mb,Baer:2006te}, while performing a scan over the constrained MSSM (CMSSM) parameter space.  Our aim is to provide an analytical understanding of this phenomenon, identifying the parameter space for which it occurs.  This will also provide a simple connection with the results of Ref.~ \cite{Cheung:BS}. In section~\ref{sec:cal}, we are going to show the suppression in SI scattering cross section due to the blind spots with an intermediate $m_A$. In section~\ref{sec:num}, we show some numerical study of the blind spots and also the scattering cross section in the blind spots mentioned in~ \cite{Cheung:BS}, which we call traditional blind spots. In section ~\ref{sec:collider}, we show how the blind spot scenario can be tested at the LHC and at future linear colliders.  We reserve
section~\ref{sec:conclusions} for our conclusions.

\section{Blind spots and the CP-odd Higgs mass}
\label{sec:cal}
\begin{figure}[tbh]{
\includegraphics[width = 5cm,clip]{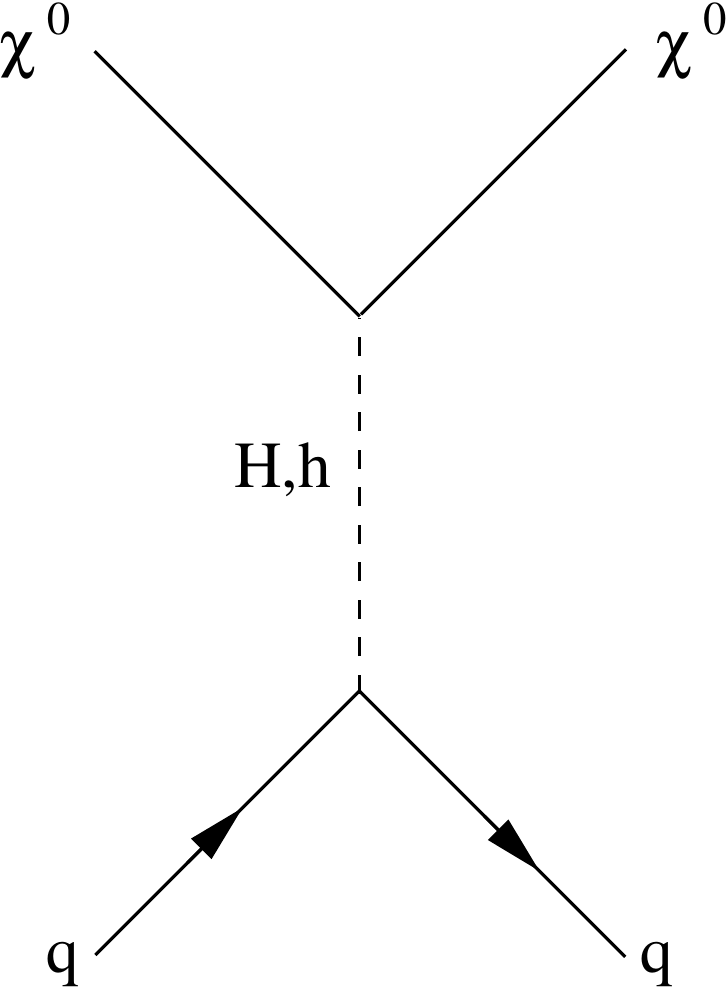}
\caption{Feynman diagram for a neutralino scattering off a heavy nucleus through a CP-even Higgs}
\label{fig:dgrm}
}
\end{figure}

In the MSSM, the SI neutralino scattering off a heavy nucleus is mediated by Higgs bosons, or squarks.  The effects from squarks are very suppressed if they are heavy, with masses larger than or of the order of a TeV. Thus, the main contributions come from the t-channel CP-even Higgs exchange, as shown in Figure\ref{fig:dgrm}.  The amplitude associated to the non-standard Higgs boson diagram is suppressed due to the relatively large values of their masses. However, this suppression may be compensated by the presence of large couplings. In particular, for moderate or large values of $\tan\beta$, the coupling of the down quarks to the heaviest CP-even Higgs boson is enhanced by $\tan\beta$ factors, while the coupling of the DM to these Higgs boson presents similar enhancements due to the larger component of  the lightest neutralino in the down-type Higgsinos at large values of $\mu$.  Therefore, the non-standard Higgs boson exchange amplitude may become of the same order as (or even larger than) the SM-like Higgs exchange one. 

Depending on the signs of the Higgsino mass parameter $\mu$ and the gaugino mass parameters $M_{1,2}$ the non-standard Higgs contribution may interfere in a constructive or destructive way with the SM-like Higgs one.  In the  case of destructive interference, for critical values of parameters, the amplitude from  light Higgs exchange and heavy Higgs exchange exactly cancel against each other, which we call generalized blind spots, since they provide a more general version of the ones previously discussed in the literature,  that are present for very large values of the non-standard Higgs masses.

First consider a neutralino scattering off a down-type quark.   As stated above, the amplitude associated with the heavy, non-standard Higgs exchange  is enhanced by $\tan\beta$.  At the tree level, the down-quarks only couples to the neutral $H_d$ component of the Higgs. The CP-even Higgs mass eigenstates can be expressed in terms of the gauge eigenstates as
\begin{eqnarray}
h &=& \frac{1}{\sqrt{2}}(\cos\alpha \ H_u - \sin\alpha \ H_d) \\
H &=& \frac{1}{\sqrt{2}}(\sin\alpha \ H_d + \cos\alpha \ H_u).
\label{eq:h}
\end{eqnarray}
Therefore,  the down-quark contribution to the SI amplitude is proportional to  
\begin{equation}
a_d \sim \frac{m_d}{\cos\beta} \left(\frac{-\sin\alpha \  g_{\chi \chi h}}{m_h^2}+\frac{\cos\alpha \  g_{\chi\chi H}}{m_H^2} \right) .
\label{eq:sig_d}
\end{equation}
Given the interactions
\begin{equation}
L \supset -\sqrt{2}g' Y_{H_{u}}\tilde{B}\tilde{H_u}H_u^{*}-\sqrt{2}g\tilde{W}^a\tilde{H_u}t^aH_u^{*}+(u \leftrightarrow d)
\label{eq:lag}
\end{equation}
and the decomposition of a neutralino mass eigenstate
\begin{equation}
\tilde{\chi} = N_{i1} \ \tilde{B} + N_{i2} \ \tilde{W} + N_{i3} \ \tilde{H}_d + N_{i4} \ \tilde{H}_u,
\label{eq:decomp}
\end{equation}
the couplings of a light or a heavy Higgs to the neutralinos are
\begin{eqnarray}
g_{\chi\chi h} &\sim&(g_1 N_{i1} - g_2 N_{i2})(-\cos\alpha \ N_{i4} -\sin\alpha \ N_{i3}) \\
g_{\chi\chi H} &\sim& (g_1 N_{i1}- g_2 N_{i2})(-\sin\alpha \ N_{i4} +\cos\alpha \ N_{i3}).
\label{eq:couplings}
\end{eqnarray} 
Then the down-quark contribution to the SI amplitude is given by
\begin{equation}
a_d \sim  \frac{m_{d}(g_1 N_{i1} -g_2 N_{i2} )}{\cos\beta}\left[N_{i4}\sin\alpha \cos\alpha \left(\frac{1}{m_h^2}-\frac{1}{m_H^2}\right) +N_{i3}\left(\frac{\sin^2\alpha}{m_h^2}+\frac{\cos^2\alpha}{m_H^2} \right) \right] 
\label{eq:sigd_med}
\end{equation}
In the above, we neglected the possible couplings of the down-quarks to the neutral components of the Higgs $H_u$, which are induced after
supersymmetry breaking.   Including them, the one loop coupling of a down quark to the neutral higgs field is given by 
\begin{equation}
L = f_d \bar{d}_L d_R H_d^0+\epsilon_d f_d \bar{d}_L d_R H_u^{0*} + h.c., 
\end{equation}
which modifies the higgs coupling to down quarks. Then Eq.~(\ref{eq:sigd_med}) becomes
\begin{dmath}
a_d \sim  \frac{\bar{m}_d(g_1 N_{i1} -g_2 N_{i2} )}{\cos\beta}\left[N_{i4}\sin\alpha \cos\alpha \left(\frac{1-\epsilon_d/\tan\alpha}{m_h^2}-\frac{1+\epsilon_d\tan\alpha}{m_H ^2}\right) + 
N_{i3}\left(\frac{\sin^2\alpha(1-\epsilon_d/\tan\alpha)}{m_h^2}+\frac{\cos^2\alpha(1+\epsilon_d\tan\alpha)}{m_H^2} \right) \right], 
\label{eq:sigd_1loop}
\end{dmath}
where $\epsilon_d \approx \frac{2 \alpha_s}{3\pi}M_3\mu C_0(m_0^2,m_R^2,|M_3|^2)$ ~\cite{Carena:2008ue}, $\bar{m}_d \equiv\frac{m_d}{1+\epsilon_d\tan\beta}$ and 
\begin{equation}
C_0(X,Y,Z) = \frac{y}{(x-y)(z-y)}\log(y/x)+\frac{z}{(x-z)(y-z)}\log(z/x).
\end{equation} 
The quantity $\epsilon_d$ is suppressed if the first and second generation squarks are much larger than the gluino mass $M_3$ and the Higgsino mass parameter $\mu$. In the following, for simplicity,  we shall assume that such a large hierarchy is present and thus in the rest of our analysis, $\epsilon_d$ 
is set to zero.  The main effect of these corrections is to modify the coupling of the heavy Higgs boson by a few tens of percent at very large values of $\tan\beta$, what leads to a small modification of the precise value of $m_H$ at which the blind spot is present. 

Following ref~\cite{pierce:decomp}, $N_{i3}$ and $N_{i4}$ are proportional to  
\begin{eqnarray}
N_{i3} &\sim& (m_{\chi}\cos\beta+\mu \sin\beta) \\
N_{i4} &\sim& (m_{\chi}\sin\beta +\mu \cos\beta). 
\label{eq:ns}
\end{eqnarray}
Also, barring the case in which $m_A$ is of the order of $m_h$,  for this analysis, we can take the decoupling limit values of $m_H$ and $\sin\alpha$, namely, $m_H \approx m_A$, and $\sin\alpha \approx -\cos\beta$. In this case,  the amplitude becomes proportional to
\begin{equation}
a_d \sim \frac{m_d}{\cos\beta}\left[\cos\beta(m_{\chi}+\mu \sin 2\beta) \ \frac{1}{m_h^2}- \mu \sin\beta \cos 2\beta \ \frac{1}{m_H^2} \right].
\label{eq:sigd_final}
\end{equation}
We can do a similar exercise for a neutralino scattering off an up-type quark, which gives
\begin{equation}
a_u\sim \frac{m_u}{\sin\beta}\left[\sin\beta(m_{\chi}+\mu \sin2\beta) \ \frac{1}{m_h^2}+ \mu \cos\beta \cos2\beta \  \frac{1}{m_H^2} \right].
\label{eq:sigp_final}
\end{equation}
Include the contributions from all quarks, including the gluon induced ones, the SI scattering cross section can be expressed as
\begin{equation}
a_p = \left(\sum\limits_{q=u,d,s}f_{Tq}^{(p)}\frac{a_q}{m_q}+ \frac{2}{27}f_{TG}^{(p)}\sum\limits_{q=c,b,t}\frac{a_q}{m_q} \right)m_p, 
\end{equation}
where $f_{Tu}^{(p)} = 0.017\pm0.008$, $f_{Td}^{(p)} = 0.028\pm0.014$, $f_{Ts}^{(p)} = 0.040\pm0.020$ and $f_{TG}^{(p)} \approx 0.91$ are the quark form factors~\cite{Junnarkar:2013ac, Hill:2011be} defined as 
\begin{equation}
<p|m_{q}q\bar{q}|p> \equiv m_{p}f_{Tq}^{(p)},  \;\;\;\;\;  f_{TG}^{(p)} = 1 - \sum f_{Tq}^{(p)} .
\end{equation}. 
Using equations (\ref{eq:sigd_final}) and (\ref{eq:sigp_final}), then the SI scattering cross section is
proportional to 
\begin{equation}
\sigma_p^{SI} \sim \left[(F_{d}^{(p)}+F_{u}^{(p)})(m_{\chi}+\mu \sin2\beta)\frac{1}{m_h^2}+\mu \tan\beta \cos2\beta(-F_{d}^{(p)}+F_u^{(p)}/\tan^2\beta)\frac{1}{m_H^2}\right]^2,
\label{eq:sig}
\end{equation}
with $F_{u}^{(p)} \equiv f_{u}^{(p)}+2\times\frac{2}{27}f_{TG}^{(p)} \approx 0.15$ and $F_{d}^{(p)} = f_{Td}^{(p)}+f_{Ts}^{(p)}+\frac{2}{27}f_{TG}^{(p)} \approx 0.14$. 
The first term denotes the contribution of the lightest Higgs and its cancellation leads to the traditional blind spot scenarios~\cite{Cheung:BS}.  The second term is  the contribution of the heavy Higgs and  as mentioned before for values of $|\mu| \simgt m_{\chi}$ and large $\tan\beta$ may become of the same  order as the SM-like Higgs one.

In the above, we have used the proton scattering amplitudes to define the spin independent scattering cross section.  The result remains valid after including the neutron contributions, since for a neutralino scattering off a neutron the form factors are $f_{Tu}^{(n)}$ = 0.011, $f_{Td}^{(n)} = 0.0273$, $f_{Ts} ^{(n)}= 0.0447$ and $f_{TG}^{(n)} = $0.917 \cite{micrOMEGA} and therefore $F_{u}^{(n)} \approx 0.15$  and $F_d^{(n)} \approx 0.14$, same as $F_u^{(p)}$ and $F_d^{(p)}$. 

Therefore, the tree-level scattering cross section due to the light  and  heavy CP-even Higgs exchange cancel against each other when 
\begin{equation}
(F_{d}^{(p)}+F_{u}^{(p)})(m_{\chi}+\mu \sin2\beta) \frac{1}{m_h^2} \simeq  F_{d}^{(p)}\  \mu \tan\beta \cos2\beta \frac{1}{m_H^2},
\label{eq:bs}
\end{equation}
which we call generalized blind spots.   Taking into account the values of $F_{u}^{(p,n)}$ and $F_{d}^{(p,n)}$ given above,  and for moderate or large values of $\tan\beta$, the blind spot can be simplified as 
\begin{equation}
2 \ (m_{\chi}+\mu \sin2\beta) \frac{1}{m_h^2}  \simeq -  \ \mu \tan\beta \frac{1}{m_H^2}
\label{eq:simple}
\end{equation}
Similar to the case in which the heavy Higgs decouples,  for intermediate values of $m_A$ the suppression due to the blind spots only happens when $\mu < 0$.  This effect was studied before~\cite{Ellis:2000ds,PhysRevD.63.065016,Baer:2006te}, and the suppression in DDMD was identified numerically from a scan of the parameter space of the CMSSM.  Our expressions provide an analytical understanding of this phenomenon.  We find out that indeed, as can be seen from Eqs.~(\ref{eq:sig})--(\ref{eq:simple}), negative values of $\mu$ have  two effects on the scattering amplitudes : On one hand, they suppress the coupling of the lightest neutralino to the lightest CP-even Higgs boson. On the other hand, they lead to a negative interference between the light and heavy Higgs exchange amplitudes.   For sufficiently low values of $m_A$ (large values of $\tan\beta$) the heavy Higgs exchange contribution may become dominant.   On the other hand,  for large values of $m_A$  the SM contribution becomes dominant and the main contribution from exchange of a heavy Higgs comes from the interference with the SM-like one and is only suppressed by $1/m_A^2$.  

 \section{numerical study}
\label{sec:num}
To perform a numerical study of the SI scattering cross section when all sfermions are heavy, the relevant parameters are the Bino mass  $M_1$,  the Wino mass $M_2$,  the  Higgsino mass $\mu$, the CP odd Higgs mass $m_A$ and $\tan\beta$. In the following, we will concentrate on the case in which LSP is mostly bino-like for simplicity, but the analysis can be easily generalized to the case in which LSP is wino-like. In the traditional blind spot scenario, at moderate or large values of $\tan\beta$ the blind spot condition, $m_{\chi} + \mu \sin 2\beta =0$, can only be satisfied if $|\mu|$ is very large, which makes the obtention of the right thermal relic density very difficult.  The generalized blind spots, instead, may be obtained for smaller values of $|\mu|$, which may be consistent  with the ones necessary to obtain a thermal DM density. 

In order to analyze the parameters  consistent with the generalized blind spots,  we first look at the parameter space away from the traditional blind spot, $\mu \sim -2 M_1$. We use ISAJET~\cite{isajet} to calculate the spectrum and the SI scattering cross section for different values of $\tan\beta$ and $m_A$, which agrees with MicrOMEGA 2.4.5~\cite{micrOMEGA} almost perfectly. We assume gaugino mass unification, so at the weak scale $M_2 \sim 2 M_1$.  As a first example, we take $M_1 \approx 220 GeV$ and $-\mu \simeq M_2 \approx 440 GeV$. 

\begin{figure}[tbh]{
\includegraphics[width = 8cm, clip]{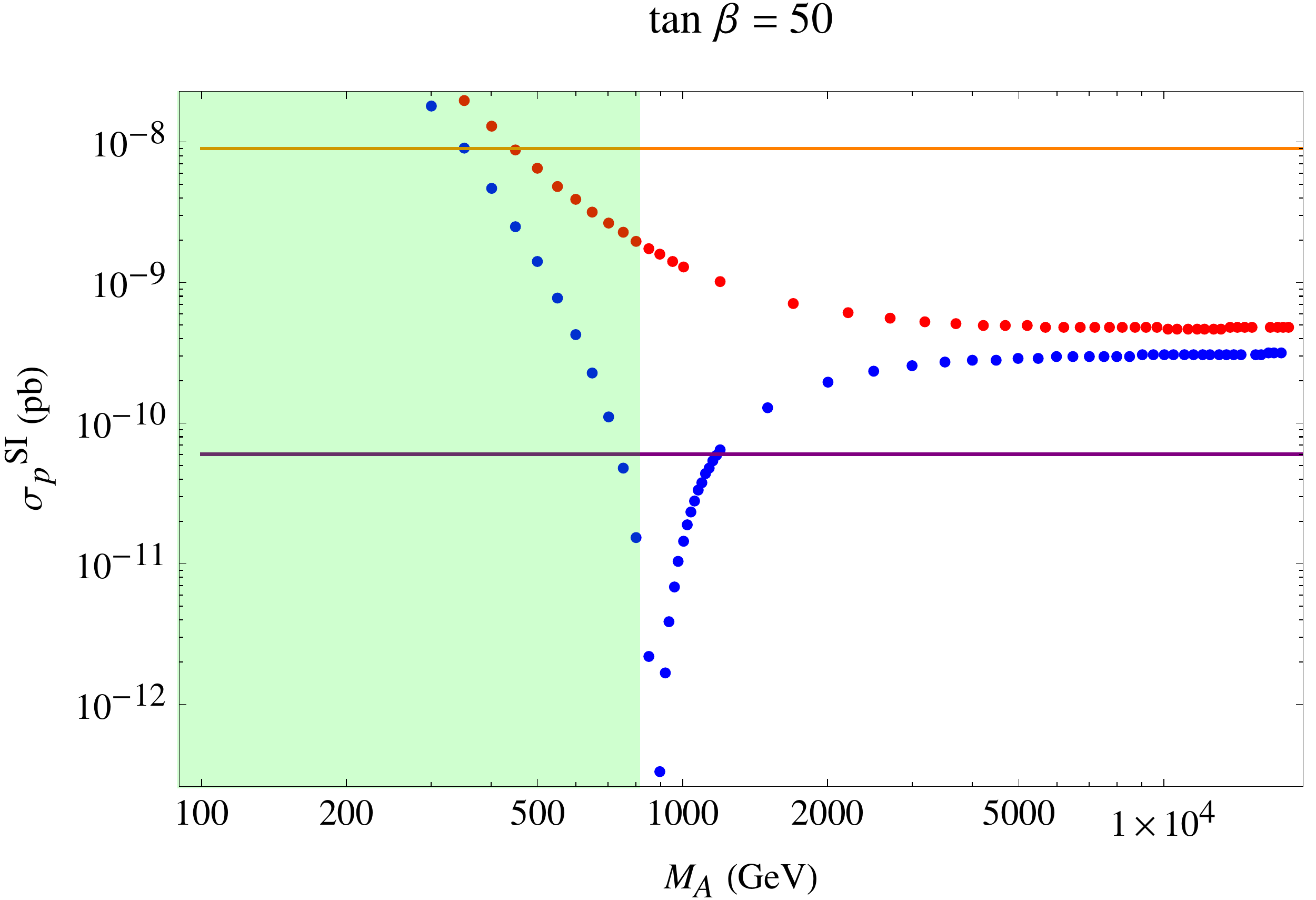}
\includegraphics[width = 8cm, clip]{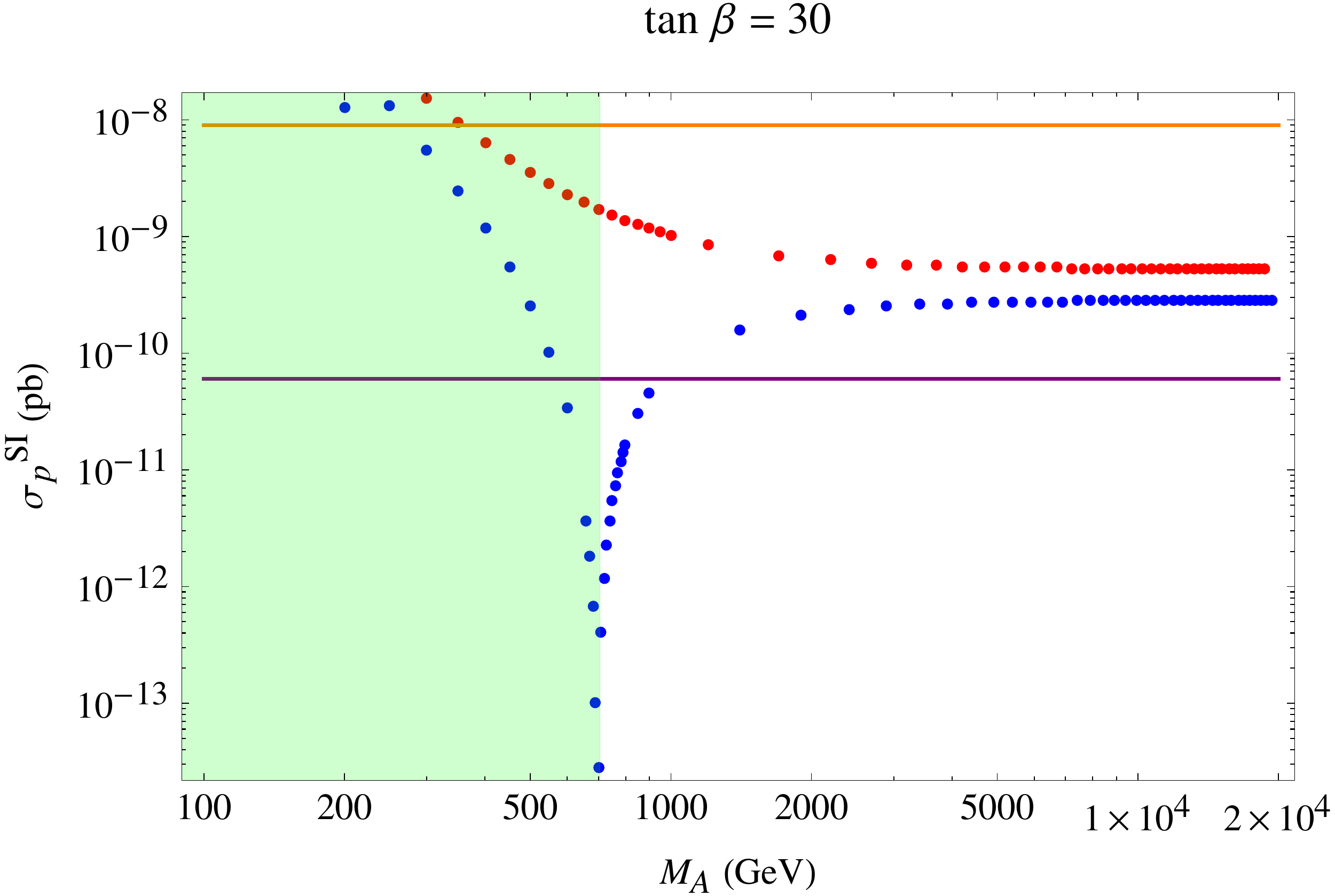}
\includegraphics[width = 8cm, clip]{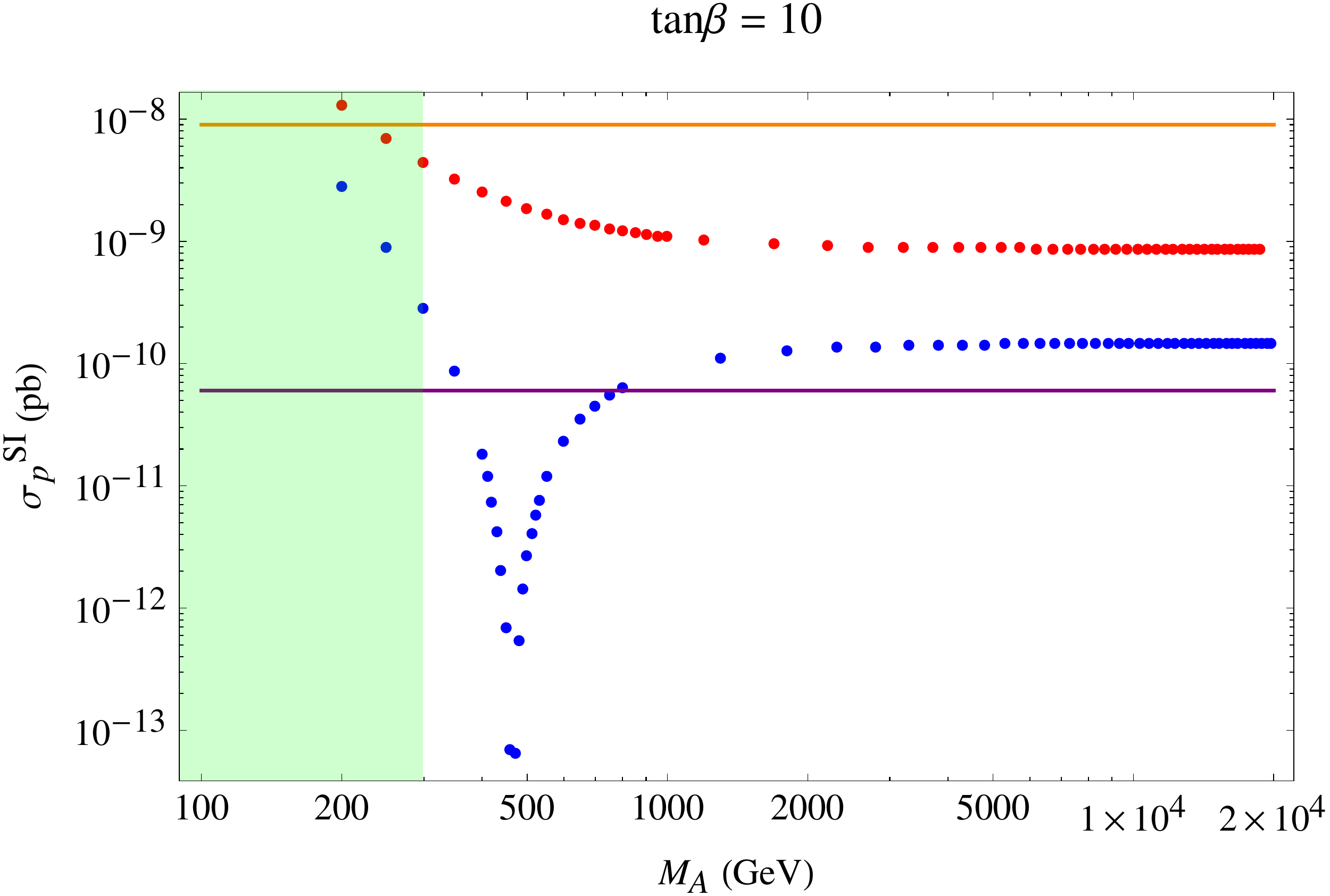}
\includegraphics[width = 8cm,clip]{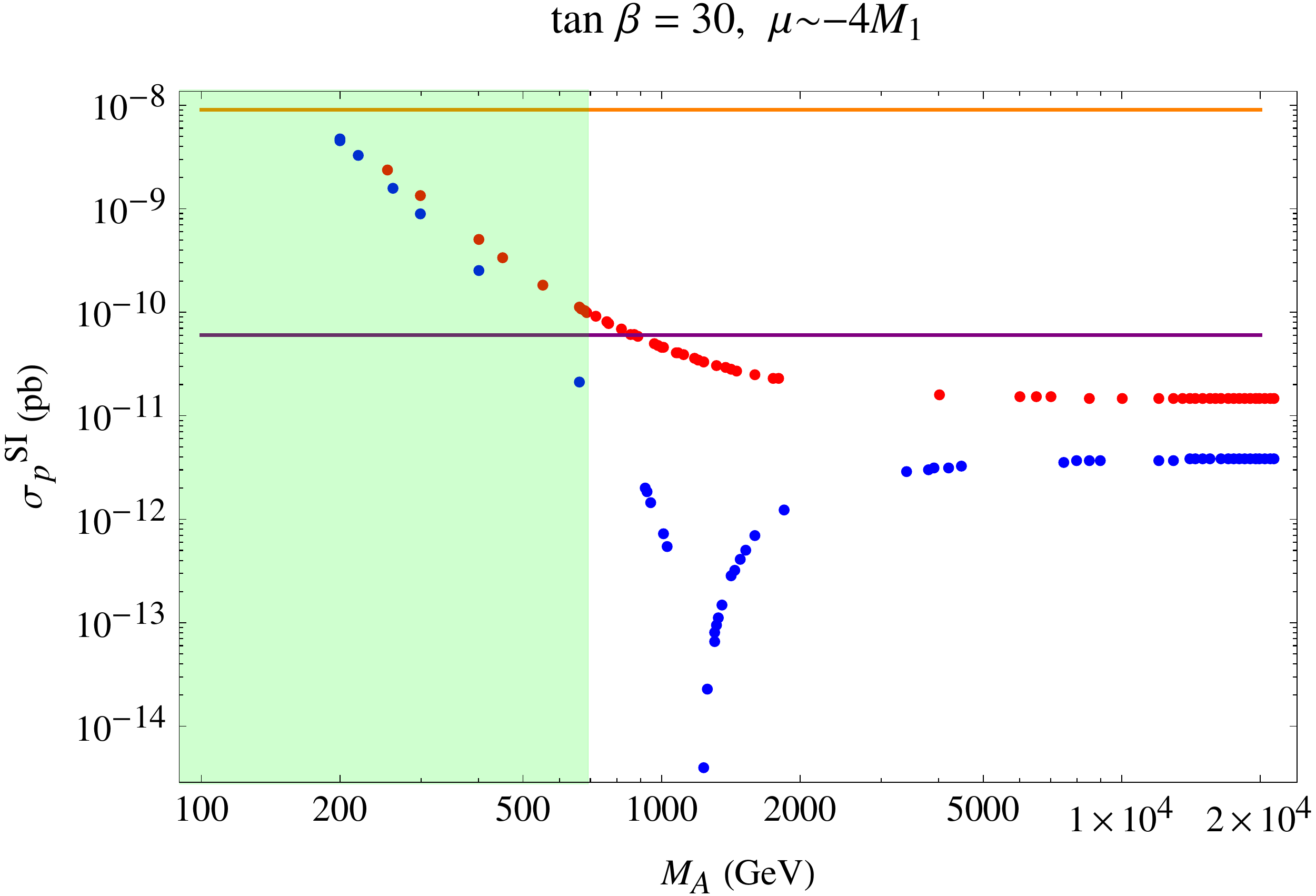}
\caption{SI scattering cross section as a function of $m_A$ for $\tan\beta = 50$~(up left), $\tan\beta = 30$~(up right) and $\tan\beta = 10$~(down left), $\mu \sim -2M_1$ and $\tan\beta = 30, \mu\sim-4 M_1$~(down right). The red dots are for the $\mu > 0$  case, and blue dots are for $\mu < 0 $ case. The green shaded area are excluded by the CMS $H,A\rightarrow \tau\tau$ searches. The orange line is the LUX limit, and the blue line is the projected Xenon 1T limit}. 
\label{fig:SI}}
\end{figure}

The SI scattering cross section as a function of $m_A$ for various values of $\tan\beta$ can be seen in Fig.~\ref{fig:SI}. For a certain value of $\tan\beta$, $m_A$ is constrained by the CMS bounds coming from  $H \rightarrow \tau \tau$ searches~\cite{CMS:htautau}, as shown in the green shaded area, where we have assumed $M_{SUSY} = 1$ TeV. The blue dots are for negative $\mu$, and the suppression due to the blind spots shown in Eq. (\ref{eq:bs}) can be seen at $m_A \approx$ 950 GeV for $\tan\beta = 50$, $m_A \approx $750 GeV for $\tan\beta = 30$ and $m_A \approx$ 500 GeV for $\tan\beta = 10$.  These values agree well with the predictions of Eq.~(\ref{eq:simple}). For comparison, in Figure~\ref{fig:SI} we also show the results for positive values $\mu = 440$~GeV (red dots), and there are no blind spots behavior as expected.  

When $m_A$ is very large, the contribution from a heavy Higgs is suppressed and the scattering cross section is  approximately  equal to  the one associated with the lightest Higgs exchange contribution. At moderate values of $\tan\beta$ even at the decoupling limit, the SI cross section is suppressed when $\mu < 0$, compared to $\mu >0$ cases.  For moderate values of $|\mu|$, the suppression is stronger for small $\tan\beta$.  Indeed,  the difference between the results for positive and negative values of $\mu$ in this regime is associated with different values of the neutralino coupling to the lightest Higgs,  $\sigma_p^{SI} \sim (m_{\chi}+\mu \sin2\beta) \frac{1}{m_h^2}$. 

In Figure~\ref{fig:SI}, we also include the LUX  limit (orange line)  for a WIMP of mass $m_\chi \simeq$~220~GeV, and the projected limit from Xenon1T (purple line). For the $M_1$ and $\mu$ values we choose here, the region allowed by the CMS $H,A \rightarrow \tau\tau$ searches is still allowed by LUX and can be probed by the future Xenon1T experiment, except for regions of parameters near the generalized blind spot. We also show a plot with $\tan\beta = 30$, and $\mu \sim -4 M_1$. In this case, the blind spot is around 1200 GeV, allowed from the CMS $H,A \rightarrow \tau\tau$ searches. 

\begin{figure}[tbh]{
\includegraphics[width = 8cm, clip]{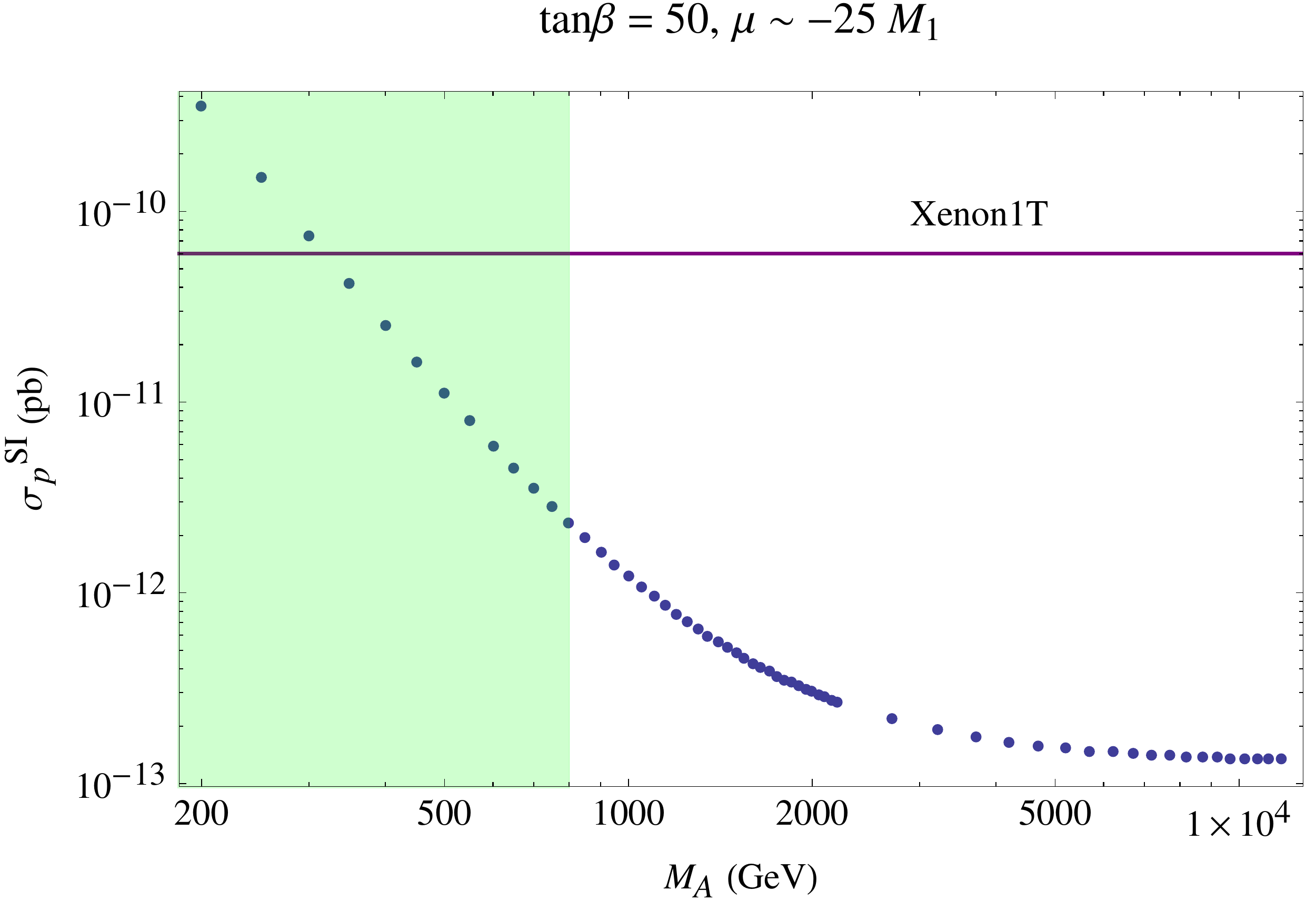}
\includegraphics[width = 8cm, clip]{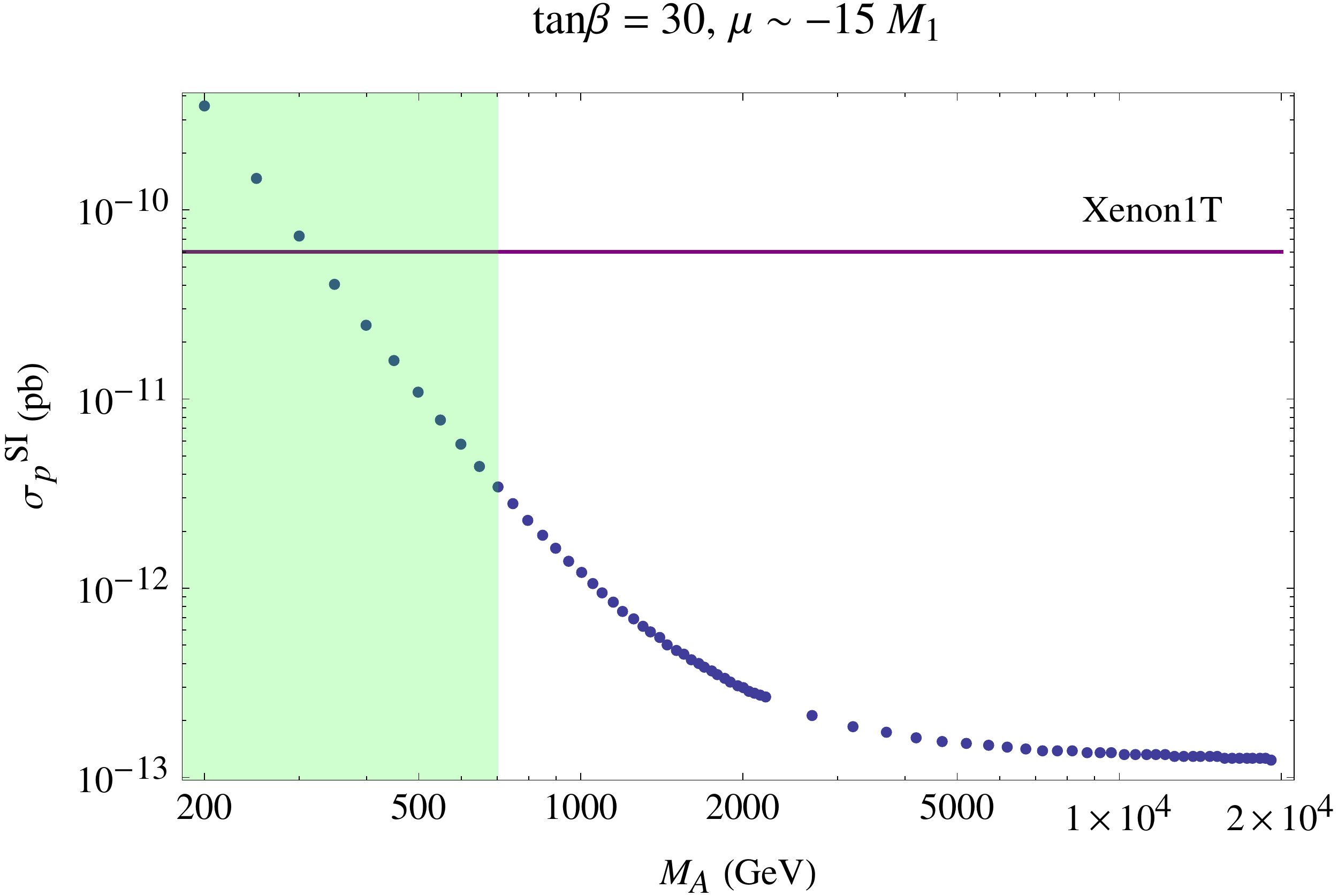}
\includegraphics[width = 8cm,clip]{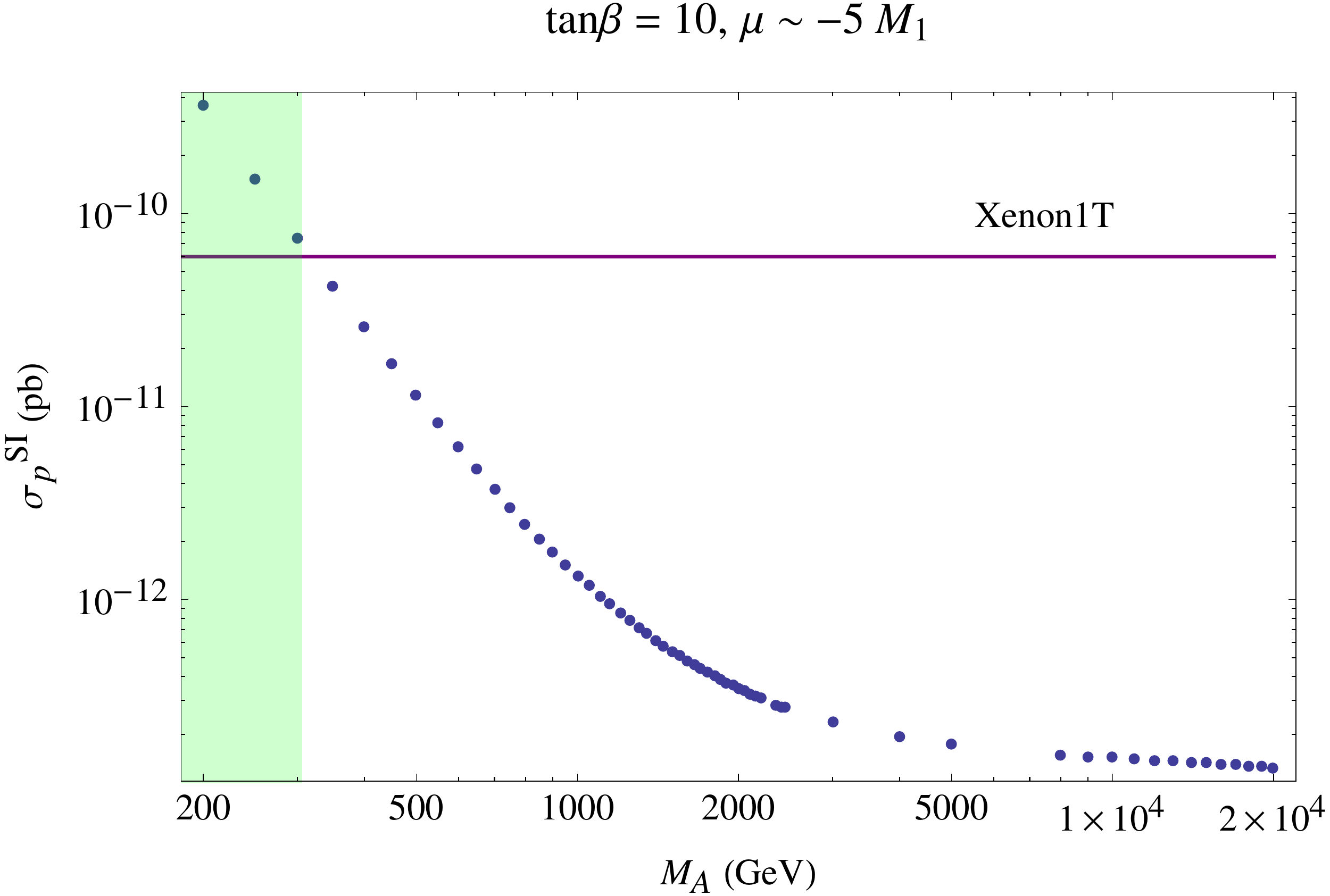}
\includegraphics[width = 8cm,clip]{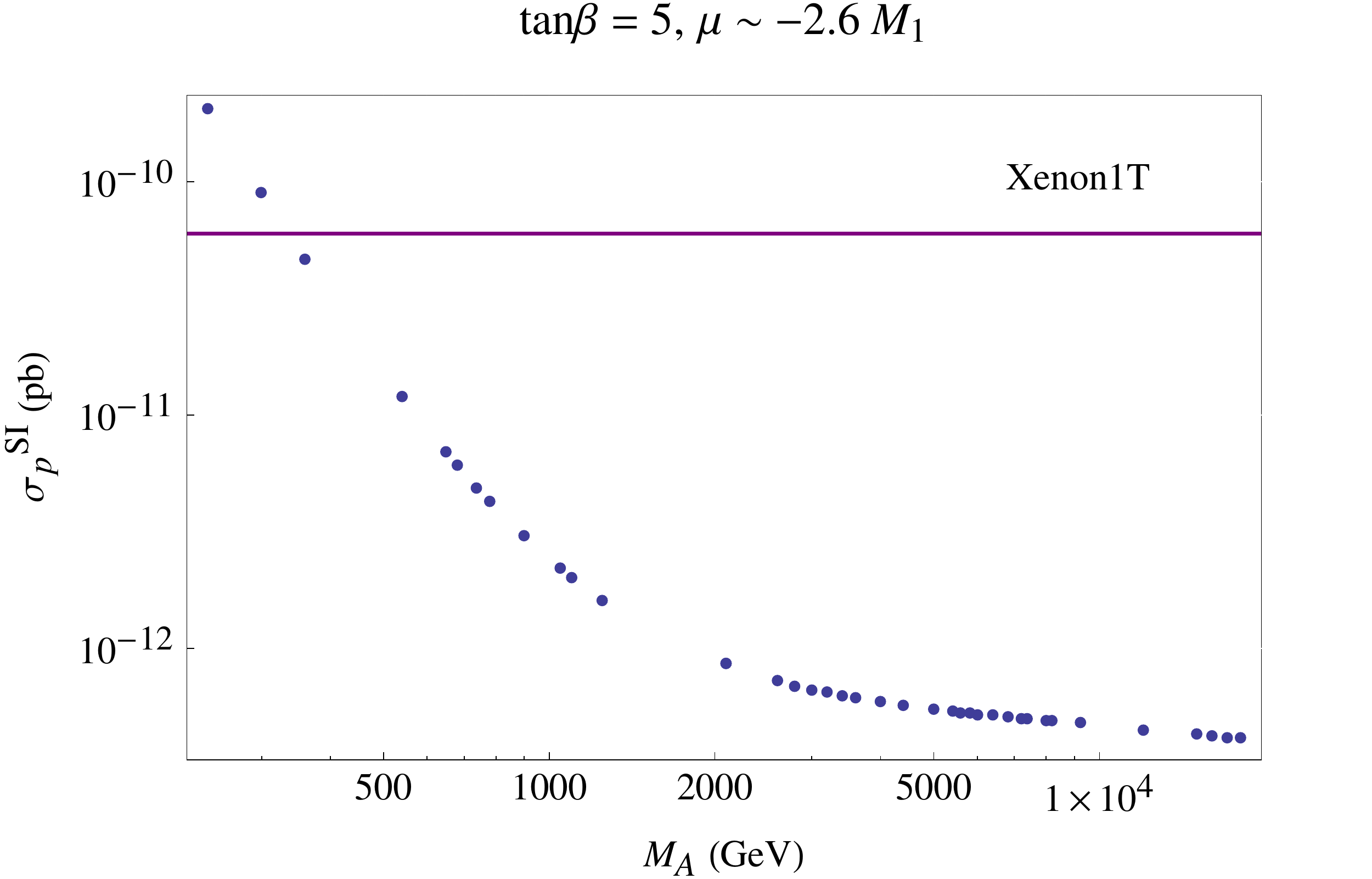}
\caption{ SI scattering cross section at the traditional blind spots where $m_{\chi}+\mu \sin2\beta = 0$ for $\tan\beta = 50$(up left), $\tan\beta = 30$~(up right), $\tan\beta = 10$~(down left) and $\tan\beta =$ 5~(down right). The green shaded area are excluded by CMS $H,A \rightarrow \tau\tau$searches.}
\label{fig:largemu}}
\end{figure}

In order to study the relevance of the generalized blind spots scenarios,  we also study what happens for moderate values of $m_A$  at the previously defined blind spot scenarios.  At the traditional blind spots where $m_{\chi}+\mu \sin2\beta=0$,  the SI scattering cross section is suppressed when $m_A$ is large, but can be sizable, and can be probed by future experiments like Xenon1T in the intermediate $m_A$ region  as shown in Figure~\ref{fig:largemu}.  For  smaller values of $\tan\beta$, like $\tan\beta =$ 5, these experiments can probe the region allowed by CMS $H,A\rightarrow \tau \tau$ searches, and can provide a probe complementary to precision $h$ couplings and future searches for non-standard Higgs bosons.  The cross section is not sensitive to $\tan\beta$, since the coupling to down fermions is enhanced by $\tan\beta$, but since $\mu$ grows together with $\tan\beta$, the down-Higgsino component is suppressed roughly by $\tan\beta$.  At large $m_A$, the cross section approaches 10$^{-13}$ pb$^{-1}$, which is below the atmospheric and diffuse supernova neutrino backgrounds. There are various contributions to this asymptotic value, including squarks, incomplete cancellation of the couplings and loop effects.  

We also analyze the relic density. Considering a thermally produced neutralino DM, the annihilation cross section is too small for Bino-like DM, which leads to DM density over abundance, while the annihilation is too efficient for pure wino or Higgsino-like DM, which results in under abundance unless the LSP is heavier than 1~TeV~\cite{Cirelli:2005uq,Cirelli:2007xd} or 2.7 TeV~\cite{Cirelli:2007xd,Hisano:2006nn}, respectively.  Only a well-tempered neutralino ~\cite{Feng:2000gh,Giudice:2004tc,Pierce:2004mk,welltempered}, a fine tuned mixture of Bino, Wino and Higgsino, can be consistent with the WMAP results $\Omega h^2 = 0.1138\pm0.0045$\cite{WMAP}.  There are two ways of annihilating excess Bino-like LSP to the correct relic density. One is to bring the mass of at least one of the sfermions down, so that there will be additional contribution from coannihilation with the light sfermion, or by the exchange of a light sfermion to provide the right relic density.  The parameter regions where this happens is often called the coannihilation and bulk regions, respectively.  Also, in the region where $M_1 \sim m_{A}/2$, which is often called A-funnel,  the LSP can annihilate resonantly into a heavy higgs, and the annihilation can be efficient enough to give the right relic density.  

\begin{figure}[tbh]{
\includegraphics[width = 8cm,clip]{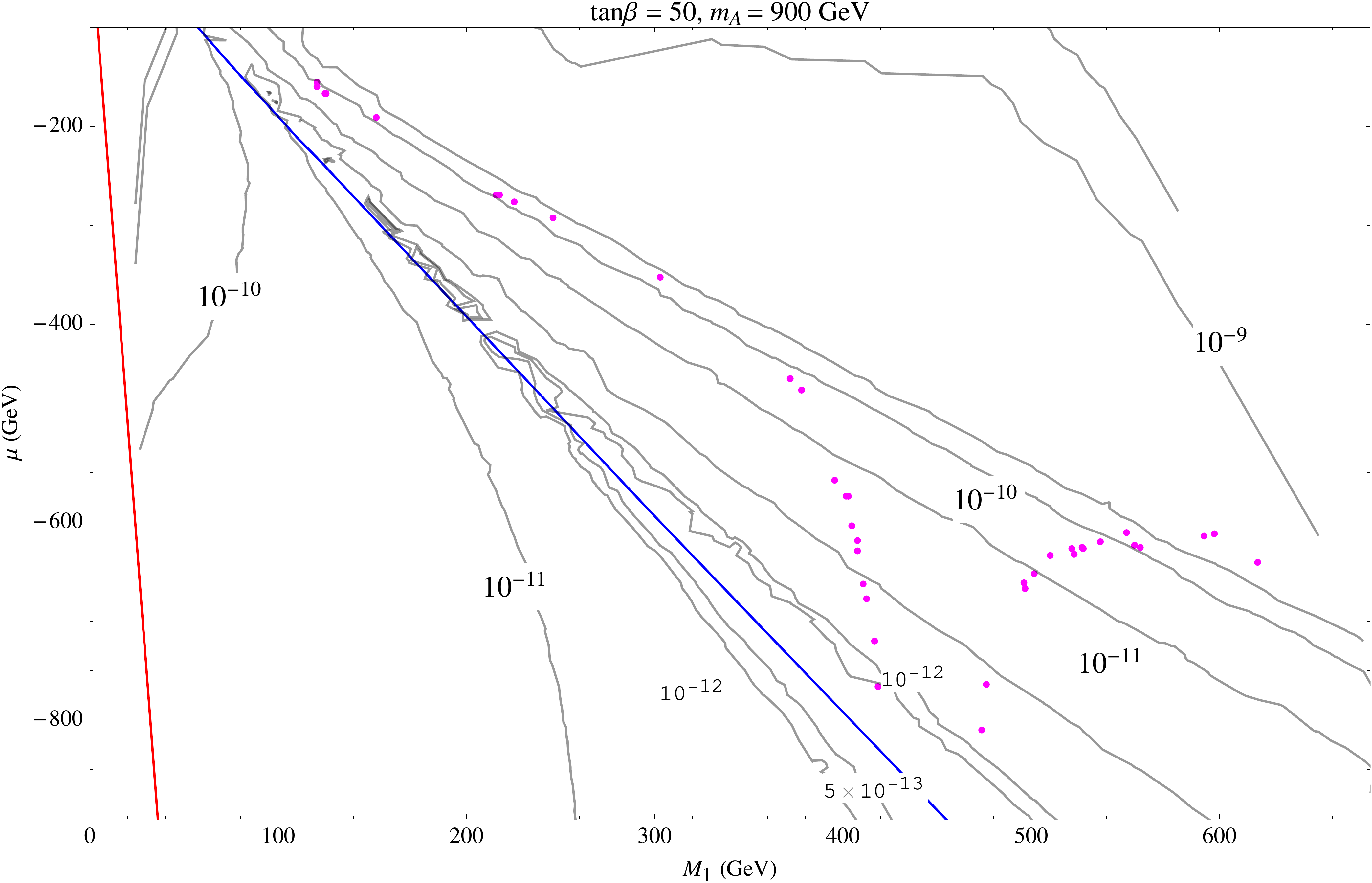}
\includegraphics[width = 8cm,clip]{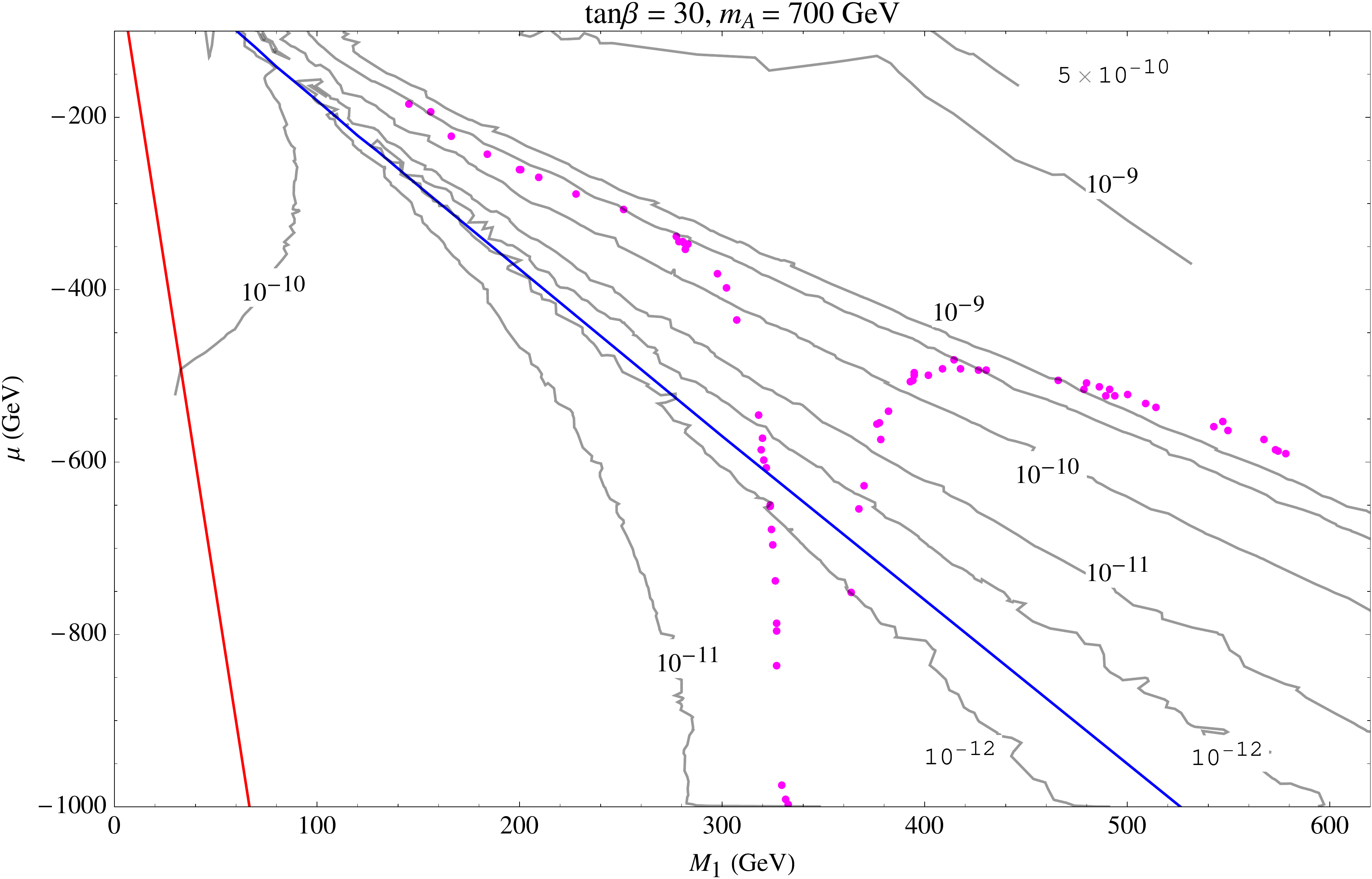}
\includegraphics[width = 8cm, clip]{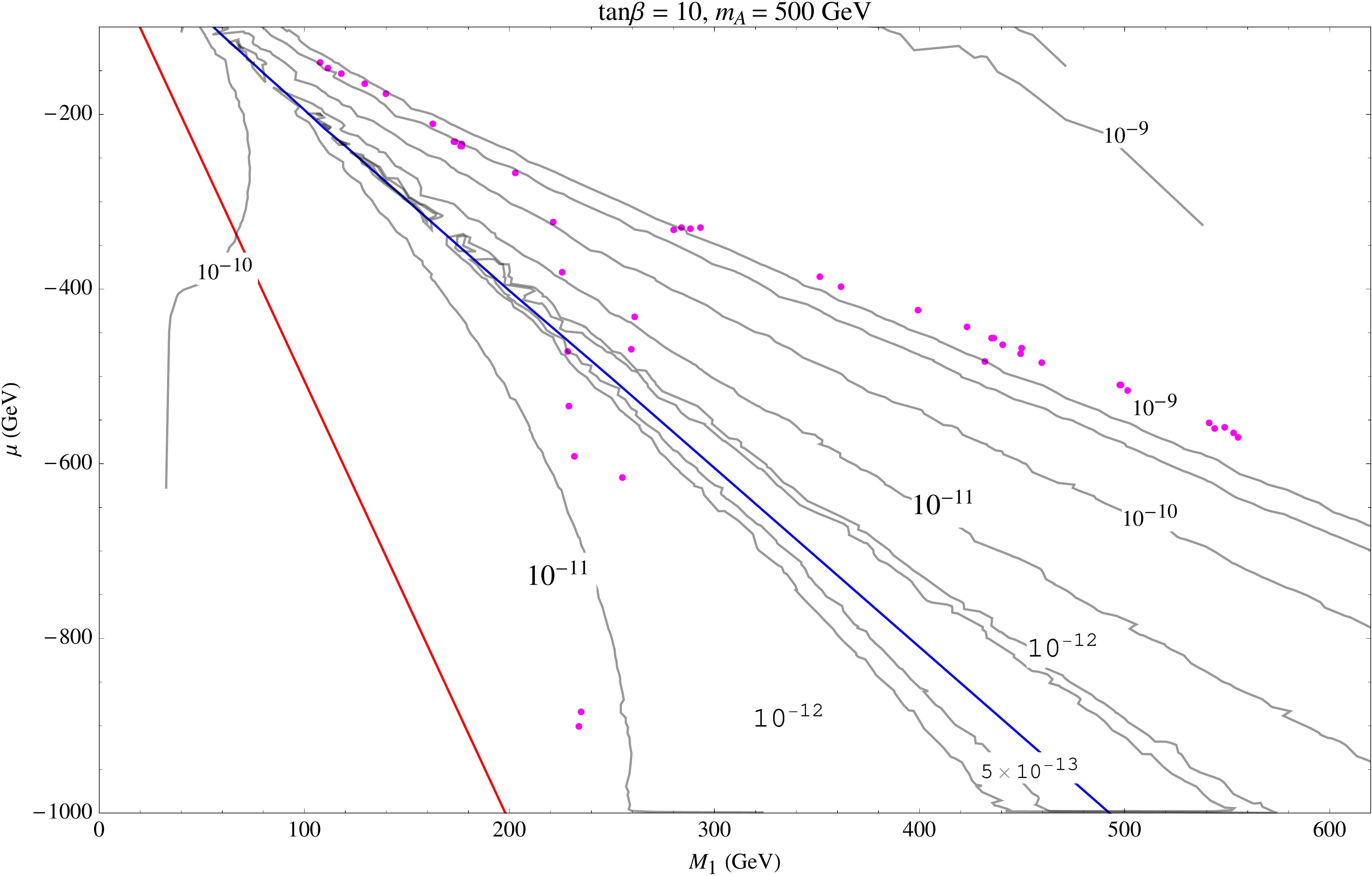}
\includegraphics[width = 8cm, clip]{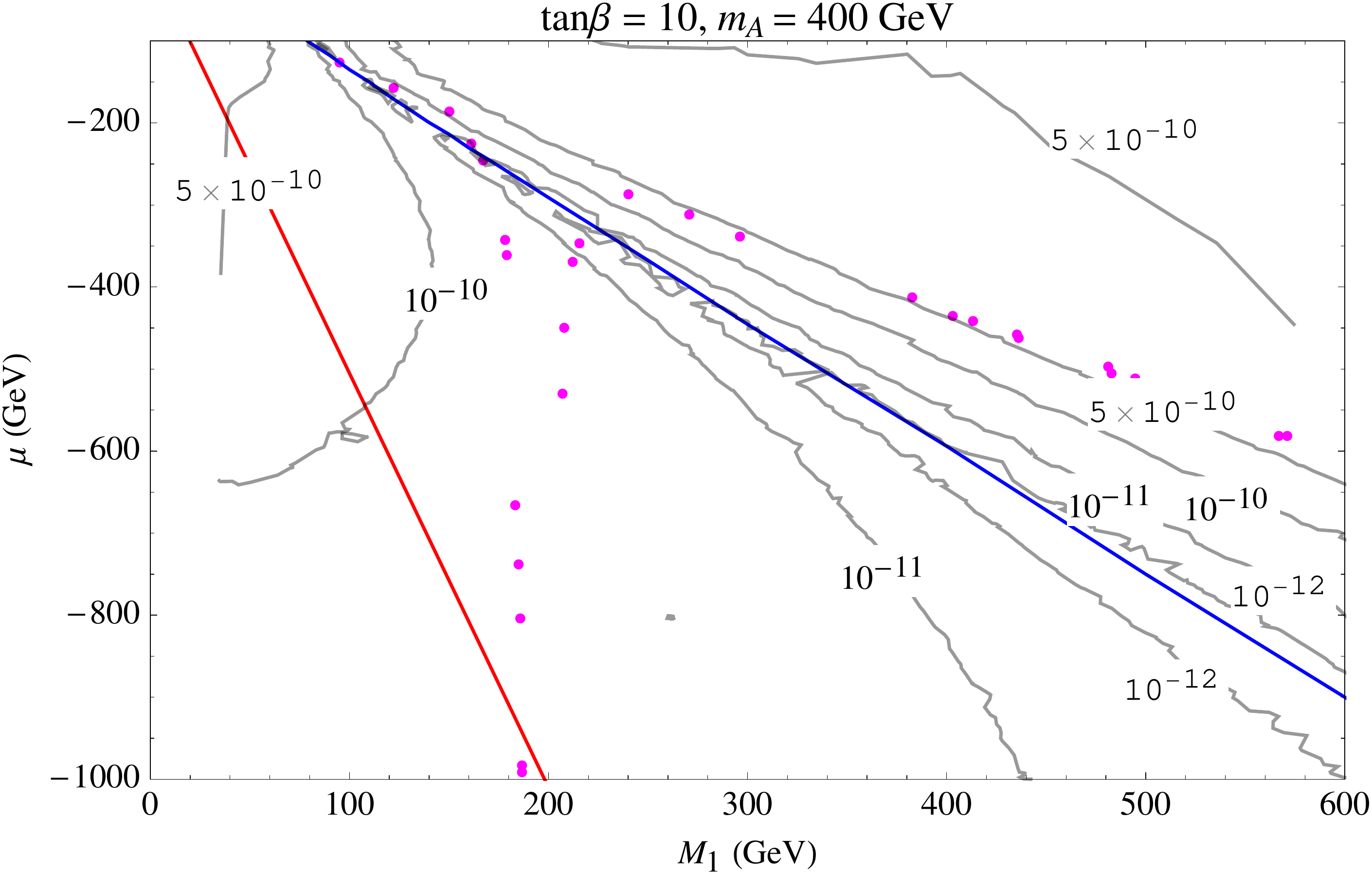}
\caption{SI DDMD cross section in the $\mu - M_1$ plane for various values of $\tan\beta$ and $m_A$, in pb. The purple dots are where the calculated relic density agrees with the WMAP value. The blue lines show the blind spot calculated from Eq,~(\ref{eq:bs}), and the red lines show the traditional blind spot where $m_{\chi} + \mu \sin2\beta = 0$.}
\label{fig:relic}}
\end{figure}

In Figure ~\ref{fig:relic}, we show the values of $M_1\simeq M_2/2$ and $\mu < 0 $ that give the right relic density for various values of $\tan\beta$.  The LEP2 experiments~\cite{LEP2} constrain the region where $M_1 < 60$~GeV, while  the region where $M_1 < 100$ GeV is also constrained by LHC trilepton searches~\cite{ATLAS:EW,CMS:EW}. The traditional blind spots fall in these experimentally constrained  regions when $\tan\beta$ is large.   For $M_1 > 100$~GeV, the blind spots from Eq. (\ref{eq:bs}) are allowed by the LHC trilepton searches in the region consistent with the dark matter relic density. The CP-odd Higgs mass $m_A$ in Figure ~\ref{fig:relic} is chosen to be consistent with the CMS $H,A \rightarrow \tau \tau$ searches. The resonant-annihilation induced by the heavy Higgs bosons can be seen from the  almost vertical lines near $m_A/2$ for $\tan\beta = $10 and $\tan\beta = $30. The tail of the resonant-annihilation region is also visible for $\tan\beta = $50.   Notice that the  region of parameters consistent with the traditional blind spots is far away with the one necessary to obtain the proper relic density. On the contrary, for values of the
CP-odd Higgs masses not much larger than the current experimental limits, for negative values of $\mu$ the region consistent with the generalized blind spots is always close to the one consistent with the observed relic density.    This means that, if $\mu$ is negative and the DM is identified with the lightest neutralino,  the SI DDMD cross section is greatly reduced to the presence of the nearby blind spots. 
 The right relic density can be obtained at the blind spot, and therefore the SI DDMD will be greatly reduced in those regions of parameters due to the cancellation of these dominant tree-level contributions.  For a given value of $\tan\beta$, when $m_A$ goes down, the blind spot moves closer to the region where a well-tempered neutralino is present. This can be seen from the $\tan\beta = 10$, $m_A = 500$~GeV and $\tan\beta = 10$, $m_A = 400$~GeV plots. In the $m_A = 400$~GeV case, the blind spot moves closer to the well-tempered region. 

\begin{figure}[tbh]{
\includegraphics[width = 12cm, clip]{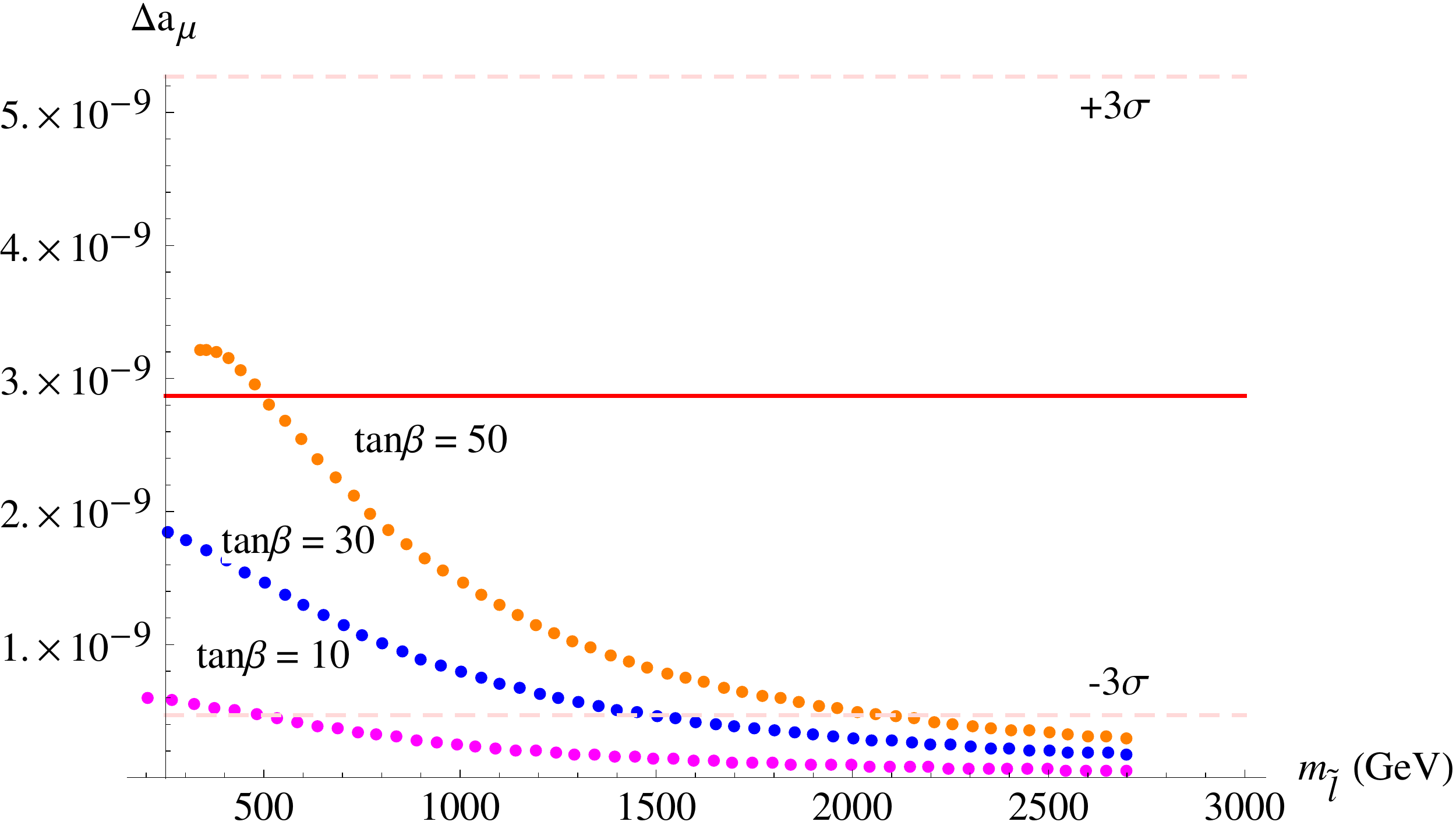}
\caption{$\Delta a_{\mu}$ at the blind spot as a function of $m{\tilde{l}}$ for $\tan\beta$ = 10(purple), 30(blue) and 50(orange). The solid red line is the central value of $\Delta a_{\mu}$, and the two dashed red lines is the $\pm3\sigma$  of $\Delta a_{\mu}$.}
\label{fig:amu}
}
\end{figure}

To be consistent with the experimental value of muon g-2 at the blind spot, where $\mu <$ 0, $M_2$ is favored to be negative~\cite{Barbieri:1982aj,Ellis:1982by,Kosower:1983yw,Moroi:1995yh,Carena:1996qa,Czarnecki:2001pv,Feng:2001tr,Martin:2001st} (for recent works see, for example, Refs.~\cite{Freitas:2014pua,Cho:2011rk}).  Fig~\ref{fig:amu} shows $\Delta a_{\mu}$ for values of the gaugino masses and the $\mu$ parameter consistent with the  generalized blind spots, Eq.~(\ref{eq:bs}), as a function of the slepton masses $m_{\tilde{l}}$ for different values of $\tan\beta$ and $M_1 = 200$~GeV, where $M_2$ and $\mu$ have been fixed at $~-2M_1$.  In this example, the value of $\Delta a_{\mu}$ is within 3$\sigma$ for $m_{\tilde{l}} < $ 800 GeV, 1.6 TeV and 2.1 TeV for $\tan\beta$ = 10, $\tan\beta$ = 30 and $\tan\beta$ = 50, respectively. The values of $\Delta a_{\mu}$ may be further enhanced if $|M_2|$ and $|\mu|$ are smaller, since the dominant contribution to muon g-2 in the MSSM is approximately proportional to 
$ \mu M_2 \tan\beta/{\rm max}(M_2^2,\mu^2,m_{\tilde{\nu}}^2)$.

Beyond the anomalous magnetic moment of the muon, there are other relevant constraints
on the supersymmetric parameter space that come from flavor and Higgs physics. The
flavor physics constraints depend strongly on the precise scale and flavor structure of the scalar mass parameters of the theory, which are not related to the direct DM detection cross section
studied in this article, and therefore we shall not discuss them further.  Similarly, the Higgs mass is obtained by pushing the stop masses to the order of a few TeV and an appropriate mixing parameter. Such heavy stops do not affect the DM searches analyzed here. The most important
effects from Higgs physics, instead, come from the Higgs couplings which are governed by tree-level processes that depend mostly on the
value of the CP-odd Higgs mass : For small values of $\mu$ compared to the stop masses, values of $m_A < 300$~GeV would induce a large mixing between the two CP-even Higgs bosons, leading to a large increase of the 
bottom-quark width and therefore to unacceptable small values of the branching ratio of the decay of the SM-like Higgs boson into gauge bosons~\cite{Carena:2001bg,Arbey:2013jla,Carena:2013ooa}.  The precise constraint on the 
CP-odd
Higgs mass depends strongly on the observed Higgs production rates. For instance,  the ATLAS
experiment sees an enahancement on both the $h \to ZZ$ and $h \to \gamma \gamma$ production rate and therefore tends to restrict values of the CP-odd Higgs mass smaller
than about 400~GeV~\cite{ATLAS-CONF-2014-010}.  In the following, we shall therefore only consider values of the CP-odd Higgs mass consistent with these bounds.

\section{Collider Tests of  the Blind Spot Scenario}
\label{sec:collider}
Eq.~(\ref{eq:simple}) provides a simple analytical expression defining the region of parameter space where the blind spots are realized at intermediate values of $m_A$. If after the next round of direct dark matter detection expreiments,  no signal is observed, it will be very relevant to understand if this is due the cancellation of the cross section amplitudes induced by the light and heavy CP-even Higgs bosons, or by other unknown effect.  In this section, we shall discuss the possibility of analyzing this question by collider searches of new Higgs bosons, charginos and neutralinos at the LHC and future lepton colliders. 

The presence of the blind spot depends on four unknown parameters, namely the mass of the heavy CP-even Higgs boson, $\tan\beta$, the  Higgsino mass parameter and the neutralino mass. 
From Eq.~(\ref{eq:simple}), for given values of $\tan\beta$ and $m_A$, the ratio of $\mu$ and $m_{\chi}$ is given by
\begin{equation}
\frac{m_{\chi}}{\mu} = - \left( \sin2\beta+\tan\beta \ \frac{ m_h^2}{2\ m_H^2} \right). 
\label{eq:simple2}
\end{equation}
For neutralino masses of the order of the weak scale, the above ratio, Eq.~(\ref{eq:simple2}), defines the Higgsino composition of the lightest neutralino, which becomes very relevant in collider searches for neutralinos and charginos.  

\begin{figure}[tbh]{
\includegraphics[width = 12cm,clip]{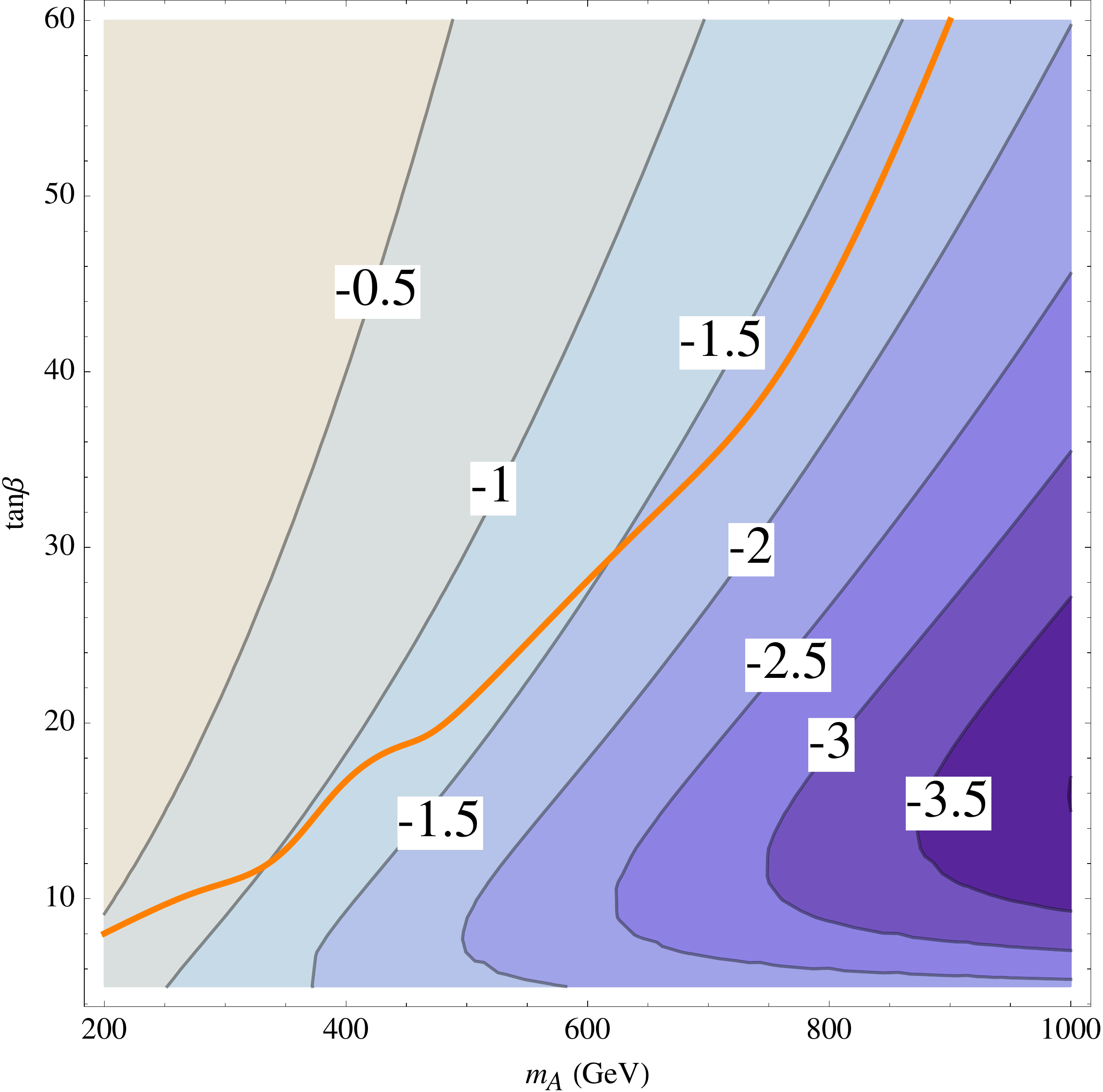}
\caption{Ratio of $\mu$ and $m_{\chi}$ at the blind spot for given values of $\tan\beta$ and $m_A$ (see Eq.~(\ref{eq:simple2})). The orange line is the current limit from CMS $H,A \rightarrow \tau \tau$ searches.}
\label{fig:ratio}
}
\end{figure}

In Fig.~\ref{fig:ratio} we show countor plots of the necessary values of the ratio $\mu/m_{\chi}$ to realize the blind pot scenario in the $m_A$--$\tan\beta$ parameter space.   We  also show in Fig.~\ref{fig:ratio} the present experimental bounds on the non-standard Higgs bosons coming from searches for $H,A \to \tau\tau$ at the CMS experiment~\cite{CMS:htautau}. Considering values of $|\mu|/m_{\chi} \geq 1$, as demanded by the requirement of $m_\chi$ being the lightest supersymmetric particle, one observes that the present experimental bounds leave open a large region of parameter space for $|\mu|/m_{\chi} > 1.5$.
Moreover,  values of the ratio of order one, consistent with the well tempered neutralino may only be obtained for smaller values of $m_A$ and $\tan\beta$. 

 Therefore, the blind spot is consistent with a well tempered neutralino only in a small, but interesting,  region of parameter space where $350$~GeV~$\simlt m_A \simlt 500$~GeV, and $6 \simlt \tan\beta \simlt 20$, where larger values of the Higgs boson masses are correlated with larger values of $\tan\beta$.  An example of such a scenario, for $m_A = 400$ GeV and $\tan\beta = 10$,  was displayed in Fig.~\ref{fig:largemu}.  
As we will discuss below, this region of parameters  will be tested by future searches for non-standard Higgs bosons in the $\Phi \to \tau\tau$ channel at the LHC.   

The LHC experiments should be able to probe more regions of parameter space for $|\mu|/m_{\chi}$ by non-standard Higgs searches.
Although there is currently no realistic estimate of the future reach of the ATLAS and CMS collaborations in the $H,A \to \tau \tau$ channel~\footnote{We thank J.~Conway, A.~Juste and J.~Qiang for  correspondence on this issue}, a conservative extrapolation would be to ignore the changes associated with the higher energy LHC run and consider a new bound by just scaling the current bound with the luminosity in such a way that cross section times the square root of the Luminosity $\cal{L}$ stays constant. At moderate and large values of $\tan\beta$ the production cross section in this channel increases as $\tan^2\beta$, so the limit for a given value of $m_A$ can be scaled in the way that $\tan^2\beta \times \sqrt{\cal L}$ stays the same. That means, at 300 fb$^{-1}$, the exclusion limit for $\tan\beta$ will be reduced roughly by a factor of $(20/300)^{1/4} \sim$~0.5. Similarly, 3~$\sigma$ evidence may be obtained at values of $\tan\beta$ of order 0.6 of the current limit. For discovery, the reach of $\tan\beta$ can be scaled by the current exclusion limit times $(20/300)^{1/4}\times (5/2)^{1/2} \sim$ 0.8. From Fig.~\ref{fig:ratio} one can see that  CMS should be able to test the blind spot scenario for values of  $|\mu|/m_{\chi}$ below 2.5, although evidence or discovery of non-standard Higgs bosons at the blind spot would demand values of $|\mu|/m_\chi$ lower than 2. We note that this estimation is quite conservative for two reasons. First, at higher energy, we expect a higher reach. Second, the cross section comes from the bottom $b\bar{b}H$ contribution, that scales like $\tan^2\beta$ and the gluon fusion contribution goes down slower than 
$\tan^2\beta$ due to the gluon fusion contribution coming from the top quark.  In summary, LHC searches for non-standard Higgs bosons  will probe the natural values of $|\mu|/m_{\chi} \simlt 2.5$,  and moderate or large values of $\tan\beta$.  On the contrary, the realization of the blind spot at large values of
$|\mu|/m_{\chi}$ will demand large values of the masses of the non-standard Higgs bosons and will be difficult to test at the LHC.

Assuming that a positive signal is observed in non-standard Higgs searches, one would acquire knowledge of two important parameters,
namely the non-standard Higgs bosons masses, and $\tan\beta$, related to the Higgs production rate.  The values of these parameters will also provide information of $\mu/m_{\chi}$, and therefore if  a well tempered thermal dark matter
may be present or resonant s-channel Dark Matter annihilation, mediated by the heavy Higgs,  is required to realize a thermal dark matter scenario. Larger values of $m_A$ and $\tan\beta$ 
will favor the latter, while smaller values will favor the former possibility.  In the latter case, information about the neutralino mass (of order $m_A/2$) would also be obtained~\footnote{In the following, we shall concentrate on the possibility of thermal dark matter, although most of the collider tests are independent of this restriction.}.

Further tests of this scenario may be obtained by direct searches for neutralinos and charginos at the high luminosity LHC.  In order to simplify our discussion, we shall assume gaugino mass unification, $M_2 \simeq 2 \ M_1$. When $M_Z< M_1 < m_h$, $\nt$ decays to a Z and the LSP with almost 100$\%$ branching ratio, and $\co$ decays to a $W^{\pm}$ and the LSP. Given the large wino pair production cross section, at 14 TeV, both ATLAS and CMS are going to reach 5$\sigma$ discovery through the tri-lepton channel for $M_1 < m_h$ ~\cite{ATLAS:projected, CMS:projected}.  Since from Higgs physics $m_A > 300$~GeV, such small values of the neutralino would not be consistent with the obtention of dark matter via resonant s-channel annihilation and therefore would demand the realization of a well tempered scenario.  Therefore, the combination of Higgs boson searches and chargino and neutralino cases will test this scenario.  The necessary presence of light winos and light Higgsinos in the trilepton signatures will provide further information of the realization of the blind spot scenario.

When $M_1 > m_h$ ($M_2 > M_1 + m_h$), $\nt$ contains an antisymmetric combination of the two higgsinos. At the same time, $m_{\no}$, which is bino-like, carries small higgsino components of opposite sign.  Considering the  matrix elements $N_{ij}$, denoting the $i$-neutralino composition on the weak eigenstate $j$ in the Bino, Wino, Higgsino-down, Higgino-up basis,  the $\nt \no Z$ coupling is proportional to $N_{13}N_{23}-N_{14}N_{24}$, so the $\nt \no Z$ coupling is suppressed compared to the $\nt \no h$ coupling. Then in the regions of $M_1 > m_h$, $\nt$ decay is dominated by the h + LSP mode.  The trilepton searches should be complemented with searches for Higgs and W bosons plus missing energy in the final state.  Unfortunately, the LHC has a limited reach in this channel, and tests of this scenario with this channel at the LHC14 will be difficult for $M_1 > 150$~GeV~\cite{CMS:projected}. 
However, for $M_1 \simeq {\cal O}( 200$~GeV),  around 15$\%$ of the higgs are boosted, so we can use the jet substructure techniques to tag the boosted higgs, which appears like a fat jet at the LHC~\cite{BDRS}.  This can provide an alternative way of testing this scenario at  LHC14, but a detailed study is lacking and is beyond the scope of this paper.  

The region of $M_1 > m_h$, and $|\mu| \simlt M_2$ can be also tested by $\co \nth$ production. $\nth$ is a symmetric combination of the two higgsinos, which leads to large $\nth \no Z$ couplings. Then $\nth$ decays to a Z and the LSP, $\nth \co$ production will give tri-lepton signatures at the LHC and CMS can probe the region where $m_{\nth} \simlt$ 600 GeV~\cite{CMS:projected}.  The kinematics and the associated production rate will provide information of the values of $\mu$ and therefore may provide a further test of the blind spot scenario at the LHC14.   

\begin{figure}[tbh]{
\includegraphics[width = 12cm, clip]{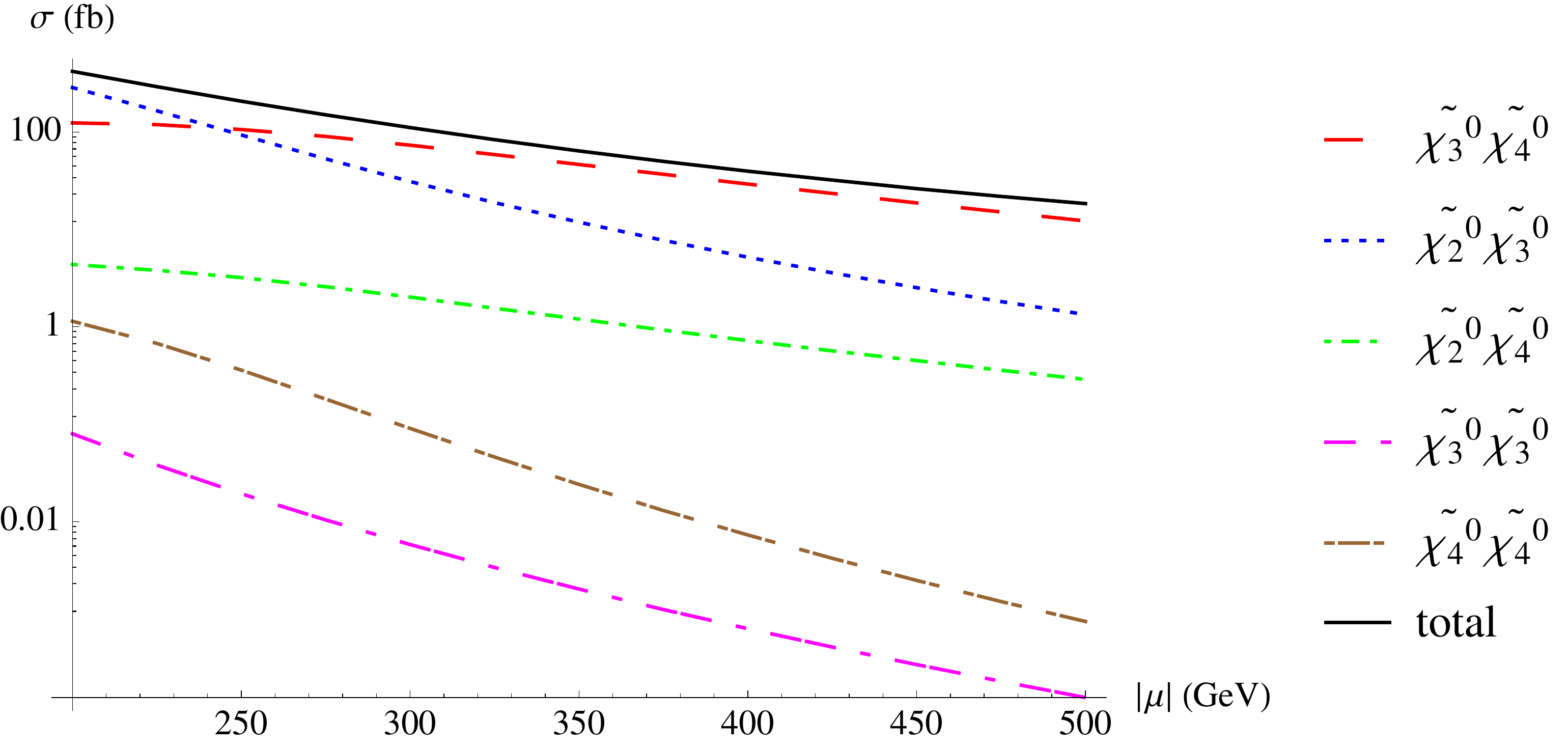}
\caption{Neutralino pair production cross section at LHC14, as a function of the absolute value of the Higgsino mass parameter $\mu$, for gaugino mass 
unification and  $M_1$ = 100 GeV.}
\label{fig:xsec_n14}
}
\end{figure}

In the above, we have concentrated on the associated production of charginos and neutralinos. 
This scenario could be further tested by neutralino pair production, which strongly depends on $\mu$ as shown in Fig~\ref{fig:xsec_n14} for $M_1 = 100$~GeV.  The $\mu$ dependence is from the higgsino components of the neutralinos and the higgsino mass. In the region where $\mu < M_2$, the production is dominated by $\nt \nth$, and the main signatures will be two $Z$-bosons plus missing energy.  Although experimental projections for  the 14~TeV LHC (LHC14) are not available, some information may be obtained from current searches at the LHC. For instance,  ATLAS has performed searches in the four-lepton final state channel~\cite{ATLAS:4lep}, and the results of this analysis do not lead to  any relevant constraint in the neutralino parameter space in this channel at the 8~TeV LHC.  However, the present bounds become comparable with the neutralino production cross sections for small values of $|\mu|$, but a few tens of GeV larger than $M_Z + m_\chi$.  It is therefore expected that LHC14 could provide information in this channel
for small values of $M_1$ and values of $\mu$  that are somewhat above the decay threshold into a $Z$ and the lighter neutralino. In the region of large $|\mu|$, $|\mu| > M_2$, $\nth \nf$ production takes over. The W + chargino mode for $\nth$ and $\nf$ opens up in this region, and is the dominant one.  Therefore, neutralinos with a dominant Higgsino components may be  search for in the four $W$'s plus missing energy mode at the LHC14. 

In the above, we have concentrated on LHC14 searches, since they provide the more immediate checks of the blind spot scenario. 
Depending on the neutralino mass, and  if $\mu$ is not too large, the value of $\mu$ can be easily probed by a linear 
collider~\cite{Baer:2011ec,Baer:2014yta}.  Depending on the center of mass energy of the linear collider, the value of $\mu$ can be either tested from neutralino pair production through the multi-lepton channel, where the neutralinos decay to a leptonic Z and the LSP, or from the $\no \nt$ production through the dilepton channel. As pointed out in Ref.~\cite{Baer:2014yta}, for the higgsino pair production with $|\mu| \sim$ 150 GeV, a signal can be observed within a few fb$^{-1}$, even in the most compressed regions of parameter space. So it is expected that the a linear collider, with a center of mass energy above the production threshold and an integrated luminosity around several fb$^{-1}$ level, will  probe large regions of parameter space consistent with the blind spot scenarios.

\section{conclusions}
\label{sec:conclusions}
The MSSM, with squark masses of the order of 1 TeV and gaugino and Higgsino masses of the order of the weak scale is an attractive scenario, that is consistent with the observed Higgs mass and contains a Dark Matter candidate, namely the lightest neutralino.  This scenario can be probed by precision measurement of the SM-like Higgs couplings, direct searches for sparticles and non-standard Higgs bosons at the LHC, as well as direct and indirect DM detection.  Future SI DDMD experiments are going to probe in an efficient way the parameter space consistent with this model and therefore it is interesting to determine the parametric dependence of the neutralino-nucleon scattering signal. 

It is well known that the SI DDMD cross section becomes smaller for negative values of $\mu$. Previous studies have also determine the presence of blind spots, where the SM-like Higgs DDMD ampltude vanishes.  In this article, we have analyzed the condition of cancellation of the SI DDMD cross section including the contribution of the non-standard Higgs bosons in the MSSM. We have shown that quite generally, this condition requires negative values of $\mu $ and we have presented analytical formulae to determine when a blind spot occurs in the MSSM.  For moderate or large values of $\tan\beta$,  and values of $m_A$ not much larger than the current limits on this quantity coming from direct searches for non-standard Higgs bosons,  the generalized blind spot scenario may occur at values of $|\mu/M_{1,2}|$  of order one,  which can lead to relic densities consistent with the observed ones.  Therefore, the generalized blind spots may become very relevant for particle physics phenomenology.  

The blind spot condition depends on parameters in the Higgs, chargino and neutralino sector. We have shown how to use the analytical understanding
provided by the blind spot condition to test this scenario. In particular, this scenario will be tested by searches for non-standard Higgs bosons, which may
provide information about the values of $m_A$ and $\tan\beta$, and therefore of $\mu/m_\chi$ in the blind spot scenario.  We have shown that this is possible,
for smaller values of $|\mu|/m_\chi < 2.5$, but becomes difficult for larger values of $|\mu|/m_\chi$.   The values of $\mu/m_\chi$ may
be further tested by chargino and neutralino searches, which complemented with the assumption of a thermal Dark Matter density, can provide an
efficient test of the realization of this scenario.  In particular, blind spot scenarios with values of $M_1 < m_h$ may be fully tested at the LHC.

\section{acknowledgments}
We would like to thank H. Baer, J. Conway, R. Hill, A. Juste, J. Qian and M. Solon for useful discussions. Work at ANL is supported in part by the U.S. Department of Energy under Contract No. DE-AC02-06CH11357. 
\bibliographystyle{utphys}
\bibliography{bsrefs}
\end{document}